# 1/*f* noise in graphene

**B Pellegrini**

Dipartimento di Ingegneria dell'Informazione, Università di Pisa, Via G. Caruso 16, Pisa 56122, Italy

E-mail: b.pellegrini@iet.unipi.it

**Abstract.** We present a novel and comprehensive model of 1/*f* noise in nanoscale graphene devices that accounts for the unusual and so far unexplained experimental characteristics. We find that the noise power spectral density versus carrier concentration of single-layer sheet devices has a behavior characterized by a shape going from the M to the Λ type as the material inhomogeneity increases, whereas the shape becomes of V type in bilayer sheet devices for any inhomogeneity, or of M type at high carrier concentration. In single-layer nanoribbons, instead, the ratio of noise to resistance versus the latter quantity is approximately constant, whereas in the bilayer case it exhibits a linear decrease on a logarithmic scale as resistance increases and its limit for zero resistance equals the single-layer value. Noise at the Dirac point is much greater in single-layer than in bilayer devices and it increases with temperature. The origin of 1/*f* noise is attributed to the traps in the device and to their relaxation time dispersion. The coupling of trap charge fluctuations with the electrode current is computed according to the electrokinematics theorem, by taking into account their opposite effects on electrons and holes as well as the device inhomogeneities. The results agree well with experiments.

1. **Introduction**

Since the discovery of graphene [1,2], its peculiar properties have been exploited in many electron applications, such as sensors, for which optimization of the signal-to-noise-ratio is an essential requirement. From this point of view, the main problem is represented by 1/*f* (or flicker) noise, whose power spectral density is inversely proportional to the frequency *f* and to the device surface *A*. As a result of possible up-conversion, flicker noise affects also high-frequency applications. Such a technological and scientific interest has motivated a number of experimental studies on 1/*f* noise in electronic graphene nanodevices [3-14]. Their flicker noise characteristics are quite unusual in comparison to those of devices based on other materials, in particular in terms of the behavior of the noise power spectral density (*PSD*) versus carrier concentration, which can exhibit a Λ [3-5,11], M [5-8,11] or V-like shape [12] for single-layer graphene sheets (SLGS), whereas most of the experimental results for bilayer graphene sheets (BLGS) [4-7,9-11,14] yield a V shape, that becomes of M type at high carrier density [5,7]. In single-layer graphene nanoribbons (SLGR) the ratio of the *PSD* to the resistance versus the latter quantity is approximately constant, whereas in bilayer nanoribbons (BLGR) it exhibits a linear decrease, if represented on a logarithmic scale, as resistance increases. Furthermore its asymptotic limit for zero resistance equals the single-layer value, whereas in the Dirac point the *PSD* is much greater for monolayer ribbons than for bilayer ones [13]. In addition, the experimental results show that the *PSD* increases with temperature [3,5,8,12].

In spite of the rich variety of experimental data, there is no unitary theory of the noise properties of graphene that is widely accepted by the scientific community. The object of the present work is to propose a novel and comprehensive model of 1/*f* noise for nanoscale graphene devices. As a corollary for other materials, we also explain the traditional empirical expression of the 1/*f* noise spectrum.

The origin of flicker noise is attributed to charge fluctuations in the traps of the device, which modulate the current flow and whose relaxation time $\tau$ is characterized by an arbitrary dispersion over a wide interval [15]. The coupling between the charge fluctuation in a single trap and the current at the device terminals is computed by means of an extension of the electrokinematics theorem [16,17] to 2D nanodevices. We consider bipolar conduction, i.e. the contribution of the currents due to both electrons and holes, on which trap charge fluctuations act in opposite ways, up to the point of canceling any effect on the total current and on the relevant noise at the Dirac point. While summing the contributions of the traps of device areas with different relative position of the Fermi level with respect to the Dirac point (resulting from any material inhomogeneities), we take into account their relative weight. In order to compute the difference $n = n_n - n_p$ (determining the charge) and the sum $n_c = n_n + n_p$ (determining the current) of the electron ($n_n$) and hole ($n_p$) surface densities, we take into account the energy dispersion of both single-layer and bilayer graphene devices and their sheet or ribbon shape, as well as the metallic, semiconducting, or mixed behaviors in the ribbons, for which we consider armchair boundary conditions. We show that there is an increase of noise with temperature and obtain numerical results exhibiting good agreement with the available experimental data.

2. **Current fluctuations**

2.1. *Extension of the electrokinematics theorem*
The primary objective is to compute the power spectral density of the current fluctuations due to the local variations of the electric field and of the carrier density, generated by the electron trapping-detrapping processes associated with the defects, as well as those due to the carrier mobility fluctuations resulting from such processes [5,13] or, independently, for instance from the migration or modification of the scattering centers [18]. To this end, we first need to compute the coupling between such local fluctuations and the current $i$ at the device terminals. This can be achieved with an extension of the electrokinematics theorem [16,17] to 2D nanodevices that we present in the following. Let us first list a few definitions: $A_0$, $\boldsymbol{E} = -\nabla A_0$ and $\varepsilon$ are the electric potential, field, and permittivity, respectively, $\boldsymbol{J}_c$ and $\boldsymbol{J} = \boldsymbol{J}_c + \varepsilon \partial \boldsymbol{E} / \partial t$ are the conduction and total current densities, respectively, $t$ is the time, $\boldsymbol{F} = -\nabla \Phi$ is an arbitrary irrotational vector with the constraint that $\nabla \cdot (\varepsilon \boldsymbol{F}) = 0$, $\Omega$ is an arbitrary volume enclosed by the surface $S$ consisting of a part $S_E$ covered by electrodes and an uncovered part $S_R$, with $S_R \geq 0$. Indeed, from the integration of the scalar products $\boldsymbol{J} \cdot \boldsymbol{F} = (\boldsymbol{J}_c + \varepsilon \partial \boldsymbol{E} / \partial t) \cdot \boldsymbol{F}$ over $\Omega$, by taking into account the divergence theorem and the fact that $\nabla \cdot \boldsymbol{J} = 0$, we obtain the electrokinematics theorem equation [16, 17]

$$\int_S (\varepsilon \frac{\partial A_0}{\partial t} \boldsymbol{F} - \Phi \boldsymbol{J}) \cdot d\boldsymbol{S} = \int_\Omega \boldsymbol{J}_c \cdot \boldsymbol{F} d^3\boldsymbol{r} \ . \tag{1}$$

A component of $A_0$ is due to the electrode potentials − which do not contribute to Eq. (1) when they are kept constant, as in the case in which we consider only the current fluctuations − and each component $A_{0j}$ is due to the $j$th charge carrier $q_j$ that is moving from one electrode of the device to the other during the time interval $t_j$ (transit time). Therefore, equation (1) becomes

$$-\int_{S_E} \Phi \boldsymbol{J} \cdot d\boldsymbol{S} + \sum_{j=1}^{M_c(t)} \int_{S_R} \varepsilon (\frac{\partial A_{0j}}{\partial t} \boldsymbol{F} - \Phi \frac{\partial \boldsymbol{E}_j}{\partial t}) \cdot d\boldsymbol{S} = \int_\Omega \boldsymbol{J}_c \cdot \boldsymbol{F} d^3\boldsymbol{r}, \tag{2}$$

where $\boldsymbol{E}_j = -\nabla A_{0j}$, and $M_c(t)$ is the total number of carriers in the space, inside and outside $\Omega$, at time $t$, and the second term becomes null in several cases of interest for the applications (Appendix A), so that the equation (2) becomes

$$-\int_{S_E} \Phi \boldsymbol{J} \cdot d\boldsymbol{S} = \int_\Omega \boldsymbol{J}_c \cdot \boldsymbol{F} d^3\boldsymbol{r} \ . \tag{3}$$

With respect to the previous results [16,17], the relationship (3) adds a new and basic equation for the study of the fluctuations in electron devices, in particular in 2D and nanoscale ones. It cannot be obtained from the Ramo-Shockley theorem [19,20], which requires that the electrodes completely enclose the device (i.e., $S_R = 0$) and which is a particular case of the electrokinematics theorem, if $\boldsymbol{F}$ becomes the electric field due to a unitary potential on one electrode and zero potentials on all the other electrodes.

If we choose $\boldsymbol{F}$ in such a way that it is $\Phi = 1$ over an electrode of surface $S_d$, threaded by the current $i \equiv -\int_{S_d} \boldsymbol{J} \cdot d\boldsymbol{S}$ of interest, and $\Phi = 0$ over all the other electrodes, from (3) we get

$$i = \int_\Omega \boldsymbol{J}_c \cdot \boldsymbol{F} d^3\boldsymbol{r}, \tag{4}$$

which is the basis of our noise calculation.

2.2. *Current*

To compute $i$ with the help of (4), we refer to a graphene field effect device with width $W$, drain-source distance $L$, and thickness $c_0$ (along the $y$, $x$ and $z$ axes, respectively); the voltage $V_g$ applied to the back gate and, possibly the voltage $V_t$ applied to a top gate, control the carrier density $n$. For the sake of simplicity, let us assume the mobilities $\mu$ of electrons and holes to be equal and independent of field and carrier densities. (The more complete case of different nobilities is discussed in Appendix B). Therefore, for a constant drain-source voltage and in the case of drift current density $J_{cx} = q\mu n_c E$ ( $E \equiv E_x$ ), that is by neglecting the velocity fluctuations of thermal origin, and by choosing $\boldsymbol{F} \equiv F_x \boldsymbol{i} \equiv -(1/L)\boldsymbol{i}$, from (4) we finally get the drain current

$$i = -\frac{1}{L}\int_A q\mu n_c E dx dy, \tag{5}$$

in which $A=LW$ and $q$ is the electron charge. Furthermore, irrespective of (5), the steady state current in any section $x$ is given by $\bar{i} \equiv I = -\int_{-W/2}^{W/2} q\mu n_c E dy = -q\mu n_c E W$, where we use the same symbols for both the time averaged and the instantaneous quantities and, for the last term, we assume that $(\mu n_c E)$ is independent (or nearly independent) of *y*.

2.3. *Fluctuations*

The aim is to evaluate the fluctuations of the current *i* that from (5) and $I = -q\mu n_c EW$ become

$$\frac{\Delta i}{I} = \frac{1}{A}[\int_A (\frac{\Delta n_c}{n_c})dxdy + \int_A (\frac{\Delta E}{E})dxdy + \int_A (\frac{\Delta \mu}{\mu})dxdy], \qquad (6)$$

where only the fluctuation terms are time dependent. The mobility fluctuations can be due to the motion or to the change of status of scattering centers that can lead to 1/*f* noise, as in the case of mesoscopic devices consisting of disordered metal or metallic glasses, as discussed in ref 18, in which the quantum mechanical variations of the conductance in the subzones of the devices could be classically described as mobility fluctuations. This could happen also in graphene nanodevices. However, since the mobility contribution to (6) does not directly depend on carrier concentration on which, on the contrary, graphene 1/*f* noise has been experimentally shown to have a strong and complex dependence, we neglect it. Therefore we attribute the origin of 1/*f* noise to the trapping-detrapping processes in defects that contribute to $\Delta i$ through the other two terms of (6), the former of which is indeed inversely proportional to $n_c$ (and, in addition, could also contribute trough possible mobility fluctuations correlated to those of the charge in the defects and of the corresponding $\Delta n_c$ [5,13]). To this end let us consider the fluctuation of the electron number $\chi = 0$, 1 in the energy level $\varepsilon_t$ of a single trap located at $r_t$ in the graphene channel or in its neighborhood. Indeed the charge fluctuation $-q\Delta\chi$ in the trap generates variations of $n_n$, $n_p$ and *E*. However, the variation $\Delta E(x, y)$ does not contribute to $\Delta i$ because it is odd in *x* and *y* around $r_t$, so that (6) becomes

$$\frac{\Delta i}{I} = \frac{1}{A}\int_A \frac{\Delta n_c}{n_c} dxdy \quad . \qquad (7)$$

Therefore, indeed, the current fluctuation are not affected by the dependence of mobility on the electric field and the carrier densities.

Since $-q\Delta\chi$ determines a potential variation $\Delta A_0$ which is equivalent to a local variation $\Delta\varepsilon_f = q\Delta A_0$ of the Fermi level $\varepsilon_f$, we have the fluctuations $\Delta n_c = a_c \Delta\varepsilon_f$ and $\Delta n = a\Delta\varepsilon_f$, where $a_c \equiv \partial n_c / \partial\varepsilon_f$ and $a \equiv \partial n / \partial\varepsilon_f$. Therefore equation (7) becomes

$$\frac{\Delta i}{I} = \frac{1}{A}\int_{\delta A}(\frac{a_c}{an_c})\Delta n dx dy = -\frac{1}{A}(\frac{a_c}{an_c})\Delta\chi, \qquad (8)$$

where the reduction of the integration surface from $A$ to the much smaller $\delta A$ around $r_t$ is justified by the fact that the effects of $\Delta n$ and $\Delta E$ fade within a few multiples of a screening length, which in the graphite c-axis is of a few Angstroms [13,21], as well as on the $x\,y$ graphene plane (Appendix C). Therefore $a_c/(an_c)$ can be assumed as a constant over $\delta A$, so that, from Gauss's theorem, we obtain $\int_{\delta A}\Delta n dx dy = -\Delta\chi$ and thus the r.h.s. of equation (8), where the variation $\Delta\chi$ occurs around the average value $\bar{\chi} \equiv \varphi = \{[1+\exp[(\varepsilon_t - \varepsilon_f)/k_B T]\}^{-1}$ given by the Fermi-Dirac factor in which $k_B$ and $T$ are the Boltzmann constant and the temperature, respectively.

3. **Noise power spectral density**

In order to compute the main quantity of the model, that is the total power spectral density $S$ of the current noise, let us recall that the power spectral density $S_t$ of the fluctuation $\Delta i$ due to a single trap becomes, according to (8), $S_t/I^2 = [a_c/(Aan_c)]^2 S_\chi$, where $S_\chi = 4\varphi(1-\varphi)\tau/[1+(2\pi f\tau)^2]$ is the Lorentzian PSD of a random telegraph signal $\chi$ and $\tau$ is the trap relaxation time [22]. For any distribution of $\tau$ (except a sharply peaked one), and even for a very small number of traps with large $\tau$, the total PSD of $i$, corresponding to the sum of the PSD $S_t$ of all the $n_t A$ (statistically independent) traps of the device, becomes [15]

$$\frac{S}{I^2} = \frac{n_t B}{A}(\frac{a_c}{an_c})^2\frac{1}{f^\gamma}, \qquad (9)$$

where $0.85 < \gamma < 1.15$ down to the frequency $1/2\pi\tau_M$, $\tau_M$ being the largest $\tau$, $n_t$ the trap surface density and $B(\varepsilon_f)$ a proper coefficient.

As a first result, we obtain the proof of the empirical relation $S \propto I^2/Nf^\gamma$ ($N = An$ being the total number of carriers in the device) for unipolar conducting materials [23] characterized by trap levels $\varepsilon_t > \varepsilon_f$, for the carriers of which (e.g., for electrons) it is actually $n_c = n \propto \exp(\varepsilon_f/k_BT)$, $a_c = a$ and, from $S_\chi \propto \varphi = \exp(\varepsilon_f/k_BT)$, $B(\varepsilon_f/k_BT) \propto \exp(\varepsilon_f/k_BT)$.

Since $n_c$ has a minimum in the Dirac point, in it we have $a_c = 0$ and thus $S = 0$. Therefore we have a second new general result that contradicts the ubiquity of flicker noise, usually motivated with the unavoidability of at least a small number of defects in a device [15]. This is a result of bipolar conduction in graphene around the Dirac point, that is of the fact that charge fluctuations in the traps generate opposite fluctuations of the electron and hole currents, thereby canceling those of the total current and the relevant contribution to the noise. However, in practice it is impossible to experimentally detect this result, because of the inhomogeneity of the doping density $N_d$ and of the trap density $n_t$ [6], and as a result of structural distortions [24] that determine a variation of the relative position of the Fermi level $\varepsilon_f$ with respect to the Dirac point in different subareas. Thus the condition $\varepsilon_f = 0$ that gives $S = 0$ is reached at different $V_g$ in different regions of the device. (Henceforth, $\varepsilon_f$ and the electron [hole] energies $\varepsilon_n$ [$\varepsilon_p$] are evaluated with respect to the Dirac level, for which, as a result of the symmetry of the graphene bands around the Dirac point, it is $\varepsilon_n(p) = -\varepsilon_p(p)$ for the same magnitude $p$ of the momentum). Therefore, for an additive quantity $H(\varepsilon_f)$, such as the carrier densities and the *PSD*, one has in general to consider the average value $<H(\varepsilon_F)> = \int H(\varepsilon_f) P(\varepsilon_{f\delta}) d\varepsilon_{f\delta}$ (Appendix D) where $\varepsilon_f \equiv \varepsilon_F + \varepsilon_{f\delta}$ with $\varepsilon_F \equiv <\varepsilon_f>$ and $P(\varepsilon_{f\delta})$ is the distribution function of the spatial variation $\varepsilon_{f\delta}$ of $\varepsilon_f$ which can be assumed of gaussian type [24] with a standard deviation $\sigma_{f\delta}$ that gives a measure of the material inhomogeneity. In particular, from (9) the *PSD* becomes

$$\frac{<S(\varepsilon_F)>}{I^2} = \frac{<n_t>}{Af^\gamma} \int B(\frac{a_c}{an_c})^2 P(\varepsilon_{f\delta}) d\varepsilon_{f\delta}, \qquad (10)$$

where the integrand is a function of $\varepsilon_f$. The term $B(\varepsilon_f)$ of (10) depends in a complex and unknown way on $\varepsilon_f$ through $\varphi(\varepsilon_f)$ in $S_\chi$ and, at the first order, it could be linearly approximated with $B \approx B_0(1+\beta\varepsilon_f/k_BT)$. Equation (10) provides a first meaningful general result to be further developed in the following sections.

At the end of this section it is also worth noting that the electrokinematics theorem allows, by means of Eq. (6), to compute the coupling between the current at the device terminals and the local fluctuations of both the number and the mobility of the carriers, for any position $r_t$ of the trap or of the scatterer generating them, i.e., in the volume or on the surface, as well as in the neighborhood of the device. Thus the debate on the origin of 1/$f$ noise from the one or the other cause, correlated or not to the same generation source, loses its significance, because both are possible, and the problem of 1/$f$ noise in electrical conductors [23,25] as a volume or surface effect vanishes, too, because, once again, both are possible, even in the same device, with different relative weight.

4. **Effects of carrier concentration on charge and current density**

4.1. *Bilayer sheet devices*

In order to compute the noise *PSD* from (10), the carrier surface concentrations $n_c = n_n + n_p$ and $n = n_n - n_p$ that determine the current and charge densities, respectively, are to be computed by determining the electron and hole densities $n_n$ and $n_p$, respectively. Owing to the cylindrical symmetry of the energy band dispersion of graphene around the Dirac point, they, in both single-layer and bilayer graphene sheets, become $n_n = (2/\pi\hbar^2)\int_0^\infty \varphi(\varepsilon_n)p\,dp$ and $n_p = (2/\pi\hbar^2)\int_0^\infty [1-\varphi(\varepsilon_p)]p\,dp$, where a factor 4 takes into account spin and valley degeneracy and $\hbar$ is the reduced Planck constant. To simplify the calculations and the representation of the results, it is convenient to normalize the energies with respect to $k_BT$ and the densities with respect to the reference density $n_0 \equiv (1/\pi)(k_BT/\hbar v)^2$, $v \approx 8.0 \times 10^5$ m/s being the in-plane velocity [26]. Therefore we define $e_n \equiv \varepsilon_n/k_BT$, $e_p \equiv \varepsilon_p/k_BT$, $e_f \equiv \varepsilon_f/k_BT$ and $e \equiv vp/k_BT$, which becomes the integration variable, as well as $e_F \equiv <e_f>$, which becomes a reference parameter controlled by the gate voltage. After such normalizations and defining $m \equiv n/n_0$ and $m_c \equiv n_c/n_0$, we get

$$m = \sinh(e_f) \int_0^\infty \frac{1}{\cosh(e_f) + \cosh(e_n)} de^2 , \tag{11}$$

$$m_c = \int_0^\infty (1 - \frac{\sinh(e_n)}{\cosh(e_f) + \cosh(e_n)}) de^2 , \tag{12}$$

where for BLGS low-energy bands, we have [26]

$$e_n = -e_p = \frac{1}{k_B T} \left( \frac{\gamma_1^2}{2} + \frac{\Delta^2}{4} + (k_B T)^2 e^2 - \sqrt{\frac{\gamma_1^4}{4} + (k_B T)^2 (\gamma_1^2 + \Delta^2) e^2} \right)^{1/2} , \tag{13}$$

in which $\gamma_1 = 0.39$ eV is the graphene interplane coupling, $\Delta$ is the asymmetry between on-site energies in the layers, which is approximately proportional to the carrier concentration $n$, i.e., $\Delta \approx cn$ with a coefficient $c \approx 1.4 \times 10^{-11}$ (meV cm$^2$) [26]. It is convenient to normalize also $a$ and $a_c$ in the form $\alpha \equiv a(k_B T / n_o) = \partial m / \partial e_f$ and $\alpha_c \equiv a_c(k_B T / n_o) = \partial m_c / \partial e_f$ so that from equations (11) and (12) we immediately have

$$\alpha = \int_0^\infty \frac{1 + \cosh(e_n)\cosh(e_f)}{[\cosh(e_f) + \cosh(e_n)]^2} de^2 , \tag{14}$$

$$\alpha_c = \sinh(e_f) \int_0^\infty \frac{\sinh(e_n)}{[\cosh(e_f) + \cosh(e_n)]^2} de^2 . \tag{15}$$

After such a normalization, from (10) the *PSD* finally reads

$$\frac{<S(e_F)>}{I^2} = \frac{<n_t> B_0}{A n_0^2 f^\gamma} \int (1 + \beta e_f)(\frac{\alpha_c}{\alpha m_c})^2 P(\eta) d\eta , \tag{16}$$

where the integrand is a function of $e_f = e_F + \eta$ and the distribution function $P(\eta)$ of $\eta \equiv \varepsilon_{f\delta} / (k_B T)$ has the standard deviation $\sigma \equiv \sigma_{f\delta} / (k_B T)$. Moreover we notice that the variation $\Delta N_d$ of the doping density $N_d$ and the local electrical neutrality at $V_g = 0$ determine the local variation $\Delta n = -\Delta N_d = a \Delta \varepsilon_f$ of the carrier concentration. Therefore we have $\sigma = \sigma_d / \alpha$ as a contribution to the standard deviation $\sigma$ due to the inhomogeneity of $N_d$, $\sigma_d$ being the standard deviation of $N_d / n_0$. In particular from (15) and (16) we again obtain $S = 0$ for $e_F = 0$ and $\sigma = 0$

*4.2. Single-layer sheet devices*

The above results hold true also for SLGS by setting $\gamma_1 = 0$ and $\Delta = 0$ in (13), as required by the linear dispersion $e_n = -e_p = e$ of its band structure. On the other hand in this less complex case, from the direct computation of $n_n$ and $n_p$, we get the simpler results

$$m = 2 \int_0^\infty \ln[\frac{\exp(e_f) + \exp(e)}{\exp(-e_f) + \exp(e)}] de \ , \tag{17}$$

$$m_c = (e_f^2 + \frac{\pi^2}{3}) \ , \tag{18}$$

$$\alpha = 2\{e_f + 2\ln[1 + \exp(-e_f)]\} \ , \tag{19}$$

$$\alpha_c = 2e_f \ . \tag{20}$$

As a check, we can obtain equation (11) for $m$ (for $e_n = e$) from (17), whereas it is not simple to obtain (12), (14) and (15) from (18), (19) and (20), respectively, or vice-versa. A numerical computation clearly confirms the equivalence of the two procedures.

*4.3. Ribbon devices*

Another objective is to extend the above equations to both single-layer and bilayer graphene ribbons for which the width $W$ becomes so small that quantum confinement effects cannot be neglected [13]; this can generate an energy band gap $\varepsilon_G$ and sub-bands in SLGR, while in BLGR, besides giving rise to sub-bands, it contributes to the already existing gap. The computation for SLGR and BLGR is made complex by the graphene crystallographic orientation, the chemical termination of the edges and the irregularities of the ribbon boundaries. However, the noise comparison between carbon nanotubes and nanoribbons shows how the "uncontrolled edge states do not seem to affect 1/*f* noise in graphene nanoribons" [13]. Also for this reason, for the purpose of the present noise study, let us consider the armchair nanoribbon model [27-30].

To this end let $a_0 = 1.42 \, \text{Å}$, $(M+1)$ and $\boldsymbol{k}_{k'} = [0, \, 4\pi/(3\sqrt{3}a_0)]$ denote the carbon-carbon (C-C) bond length, the dimer number, and the Dirac point in the Brillouin zone, respectively, so that we have $W = \sqrt{3}a_0 M/2$ and $\tilde{W} \equiv \sqrt{3}a_0(M+2)/2$, whereas the allowed wave vectors in the *y* direction become $k_{yl} = l(\pi/\tilde{W})$, where *l* is an integer in the range (0, $M+2$). The energy dispersion around $\boldsymbol{k}_{k'}$ can be approximated with the Taylor series expansion

$$\varepsilon_l = \pm v\hbar\sqrt{k_x^2 + (k_{yl} - k_{k'y})^2} = \pm v\hbar\sqrt{k_x^2 + [l - 2(M+2)/3]^2(\pi/\tilde{W})^2}, \text{ where } v = 3a_0 t_b/(2\hbar) \approx$$

8.73 x $10^5$ m/s is the in-plane velocity, $t_b \approx 2.7$ eV being the C-C bonding energy. If $(2M+3)/3 \equiv l_s'$ is an integer, by setting $l = l_s' + \phi$ we get a sub-band branch $\varepsilon_l = \pm\sqrt{(vp_x)^2 + (3\phi-1)^2(v\hbar\pi/3\tilde{W})^2}$, where $\phi = 0, \pm 1, \pm 2, \ldots$ and $p_x = \hbar k_x \equiv \varepsilon_x/v$, $\varepsilon_x$ being the momentum and the energy associated to the motion in the *x* direction. Since the number series $|3\phi \mp 1|$ does not contain 3 and its multiples, in an equivalent way we have $\varepsilon_l = \pm\sqrt{\varepsilon_x^2 + \theta^2(\varepsilon_G/2)^2}$ where $\varepsilon_G \equiv 2v\hbar\pi/3\tilde{W}$ and $\theta = [6h - 3 - (-1)^h]/4$ with $h = 1, 2, \ldots$. Instead if $(2M+5)/3 \equiv l_s''$ is an integer, by setting $l = l_s'' + \phi'$ we have $\varepsilon_l = \pm\sqrt{\varepsilon_x^2 + (3\phi'+1)^2(\varepsilon_G/2)^2}$, with $\phi' = 0, \pm 1, \pm 2, \ldots$, and again the above sub-band branch type. In both cases the ribbon behaves as a semiconductor with an energy gap $\varepsilon_G$. If, finally, $(2M+4)/3 \equiv l_m$ is an integer, by setting $l = l_m + \lambda$ we get two sub-band branches $\varepsilon_l = \pm\sqrt{\varepsilon_x^2 + (3\lambda)^2(\varepsilon_G/2)^2}$, with $\lambda = 0, 1, 2, \ldots$, which lead to a metallic behavior. Therefore, more in general, we can write $\varepsilon_l \equiv \varepsilon_h = \pm\sqrt{\varepsilon_x^2 + s^2(\varepsilon_G/2)^2}$ where $s = \theta$ and $s = 3\lambda = 3(h-1)$ for the semiconducting and metallic SLGR, respectively, and $h = 1, 2, 3, \ldots$. By defining the normalized energies $e_x \equiv \varepsilon_x/k_B T$, $e_W \equiv \varepsilon_G/2k_B T$, $e_{nh} = -e_{ph} \equiv |\varepsilon_h|/k_B T$, the electron energy in the *h*th sub-band of the SLGR becomes $e_{nh} = -e_{ph} = (e_x^2 + s^2 e_W^2)^{1/2}$. If we substitute this expression in the place of *e* into equation (13), we obtain an extension of such equation to the energy dispersion around the Dirac point of the BLGR *h*th sub-band in the form

$$e_{nh} = -e_{ph} = \frac{1}{k_B T} \left( \frac{\gamma_1^2}{2} + \frac{\Delta^2}{4} + (k_B T)^2 (s^2 e_W^2 + e_x^2) - \sqrt{\frac{\gamma_1^4}{4} + (k_B T)^2 (\gamma_1^2 + \Delta^2)(s^2 e_W^2 + e_x^2)} \right)^{1/2}, (21)$$

where we have $s = [6h - 3 - (-1)^h]/4$ and $s = 3(h-1)$ for semiconducting and metallic ribbons, respectively, with $h = 1, 2, 3, \ldots$ . For $\gamma_1 = 0$ and $\Delta = 0$ equation (21) again holds true for SLGR.

In the range $d|p_x|$ there is a density of states $dn_s = [2r/(\pi W)]d|p_x/\hbar|$ that takes into account the electron spin degeneracy. For the semiconducting case $r = 1$, corresponding to the absence of a degeneracy of the energy bands, whereas for the metallic case $r = 2$, due to the two sub-band branches. Using the expressions for $e_W$ and $e_x$, we can write $dn_s = \{6rn_0 e_W M / [\pi(M+2)]\}de_x \approx (6rn_0 e_W / \pi)de_x$, for $M \gg 2$. In analogy with what we have done before for BLGR we obtain

$$m = (6r/\pi) e_W \sinh(e_f) \sum_{h=1}^{u} \int_0^\infty \frac{1}{\cosh(e_f) + \cosh(e_{nh})} de_x, \quad (22)$$

$$m_c = (6r/\pi) e_W \sum_{h=1}^{u} \int_0^\infty (1 - \frac{\sinh(e_{nh})}{\cosh(e_f) + \cosh(e_{nh})}) de_x, \quad (23)$$

$$\alpha = (6r/\pi) e_W \sum_{h=1}^{u} \int_0^\infty \frac{1 + \cosh(e_{nh})\cosh(e_f)}{[\cosh(e_f) + \cosh(e_{nh})]^2} de_x, \quad (24)$$

$$\alpha_c = (6r/\pi) e_W \sinh(e_f) \sum_{h=1}^{u} \int_0^\infty \frac{\sinh(e_{nh})}{[\cosh(e_f) + \cosh(e_{nh})]^2} de_x, \quad (25)$$

where the upper limit $u$ of the summation is such that for $h > u$ the contribution of the $h$th sub-band becomes negligible (Appendix E). Moreover we exploit the experimental value [31] $\varepsilon_G = \alpha_H /(W - W^*)$ for the energy gap, where $\alpha_H = 0.2$ eV nm, $W^* = 16$ nm and $(W - W^*)$ can be considered as the ribbon 'effective' width in place of the geometric one $W$. In addition we also note that the experimental data show an independence of $\varepsilon_G$ on the overall crystallographic direction, thereby suggesting that the "detailed edge structure plays a more important role than such a direction in determining the properties of the graphene ribbons" [31] and that lack of knowledge (and variability within the same device) of such an edge structure may justify the above choice of the armchair ribbon model for the 1/*f* noise study.

5. **Noise dependence on the temperature**

Let us show how the model can also take into account the dependence on the temperature *T* of the contribution $S_t$ to the noise *PSD* of each trap through (*i*) $1/n_0^2 \propto 1/T^4$, (*ii*) the factor $4\varphi(1-\varphi) = 1/\cosh^2[(\varepsilon_t - \varepsilon_f)/(2k_B T)]$ of $S_\chi$, (*iii*) the factor $(\alpha_c/\alpha m_c)^2$ by means of $e_f = \varepsilon_f/k_B T$ and (*iv*) $\tau/[1+(\omega\tau)^2]$ where $\tau = \tau_0 \exp(\varepsilon_a/k_B T)$ in the assumption of thermally activated charge emission from the traps, $\tau_0$ and $\varepsilon_a$ being a proper time and activation energy, respectively. Therefore, in order to verify whether $S_t$ increases or decreases as the temperature increases, let us compute its derivative with respect to *T*

$$\frac{\partial S_t}{\partial T} = \frac{S_t}{T}[(e_f - e_t)\tanh(\frac{e_f - e_t}{2}) + e_a \frac{(\omega\tau)^2 - 1}{(\omega\tau)^2 + 1} - \xi], \quad (26)$$

where $e_a \equiv \varepsilon_a/k_B T$ and

$$\xi = 4 - 2e_f(\frac{\alpha_c}{m_c} + \frac{1}{\alpha}\frac{\partial \alpha}{\partial e_f} - \frac{1}{\alpha_c}\frac{\partial \alpha_c}{\partial e_f}). \quad (27)$$

In (27) the derivative can be computed from (14) and (15) for BLGS, obtaining however complex results. Moreover, according to (11), (13), (14) and (15), in BLGS we ought to take into account also the complex temperature dependence of $m_c$, $\alpha$ and $\alpha_c$, due to the contributions $\gamma_1/(k_B T)$ and $\Delta/(k_B T)$ to $e_n$ in (13). This does not happen for SLGS, for which the derivatives can be obtained immediately from (19) and (20) in the form $(\partial \alpha/\partial e_f) = 2[\exp(e_f) - 1]/[\exp(e_f) + 1]$ and $(\partial \alpha_c/\partial e_f) = 2$. For these, in particular, we obtain the maximum value $\xi = 6$ at $e_f = 0$ and $\xi = 0$ for $|e_f| \to \infty$, in practice for $e_f > 5$, and the first term on the right hand side (r.h.s.) of (26) becomes $\approx |e_f - e_t|$. Moreover the Fermi level is the same across the whole device and the variation of its distance $\varepsilon_f$ from the Dirac point in the different subareas acts on $\xi(e_f)$, but not on the first two terms of the r.h.s. of (26), so that after performing the average of its two members, which we have to perform to get the total *PSD*, we have to substitute $S_t$ with $<S_t>$ and $\xi$ with $<\xi(e_F)>=\int [\alpha_c/(\alpha m_c)]^2 \xi P(\eta)d\eta / \int [\alpha_c/(\alpha m_c)]^2 P(\eta)d\eta < \xi(e_f)$, which, for example, yields $<\xi(0)>$ equal to 5.46, 3.75, 3.30 and 3 for $\sigma_d$ equal to 1, 5, 10 and $\infty$, respectively. Therefore in the sum of $<S_t>$ contributions that has to be performed to obtain the total *PSD*, all the traps with $|\varepsilon_f - \varepsilon_i| + \varepsilon_a\{[(\omega\tau)^2-1]/[(\omega\tau)^2+1]\} > k_B T <\xi(\varepsilon_F)>$ (being $0 \leq <\xi(e_F)> \leq 6$) give positive contributions to the derivative of the total *PSD* with respect to the temperature. In conclusion, as most of the traps can satisfy such a condition, an increase of 1/*f* noise with temperature becomes highly possible, even if in certain ranges of the gate voltage, of the temperature and of the frequency the contrary can happen for some devices. This has been experimentally observed in ref 8 between 50 and 80 K in on-substrate SLGS in which, however, noise increases with temperature between 30 and 50 K and between 80 and 300 K. The same happens in the whole range 30-300 K in suspended SLGS. Except for such a case, the other experiments [3,5,8,12], all performed on SLGS, give a *PSD* increase with temperature. This happens also for multi layer devices [3,5].

6. **Results and discussion**

6.1. *Sheet devices*

Let us now compute numerical results from the preceding model and let us compare them with the many experimental data in the literature [3-14,25]. We first evaluate the noise of SLGS and BLGS, for both of which we use, for comparison, the same fitting parameter $\sigma_d$ in the standard deviation $\sigma = \sigma_d/\alpha$ of $P(\eta)$ where $\alpha$ is evaluated at $e_f = 0$. In the case of BLGS, in order to take into account the contribution $\Delta \approx cn$ [26] to the energy gap due to the carrier density *n*, we set, with a good approximation, $n = n_0(a'e_f + b'e_f^3)$, being $a' = 18$ and $b' = 0.10$ (Appendix D). Moreover, at room temperature and for $v = 8 \times 10^5$ m/s [26], it is $n_0 = 7.7 \times 10^{10}$ cm$^{-2}$ (whereas it is $n_0 = 6.5 \times 10^{10}$ cm$^{-2}$ for the previous value $v = 8.73 \times 10^5$ m/s). For the comparison with experimental data it is worth noting that most devices are fabricated on 300 nm SiO$_2$ substrates [5-9,11-13], so that we have $<n> = \gamma_0 V_g$ with $\gamma_0 = 7.2 \times 10^{10}$ V$^{-1}$cm$^{-2}$, while it is also $<n> \equiv n_0 <m>$ with $n_0 = (6,5 \div 7,7) \times 10^{10}$ cm$^{-2}$. Therefore the plots versus $V_g$ roughly have the same abscissa scale as those versus $<m> \equiv m_a$.

In order to compare the properties of single-layer and bilayer graphene sheets, let us evaluate the normalized quantities $e_F$, $<m_c> \equiv m_{ca}$, $\alpha$ and $\alpha_c$ versus the average normalized density $m_a$ of the charge carriers, for example for $\sigma_d = 4.5$. Their plots in figure 1 show that $|e_F|$ is much smaller and $m_{ca}$ (around $m_a = 0$) and $\alpha$ are much greater for BLGS than for SLGS, and that the $m_a$ interval within which $(\alpha_c/\alpha)^2$ varies between 0 and 1 is about four times greater in the former than in the latter case. As a first consequence, as shown in figure 1c, we have a remarkable lower value (for $|m_a| < 20$) and a weaker gate dependence of the resistance $R = L/(q\mu n_0 m_c W) \propto 1/m_c$ for BLGS than for SLGS.

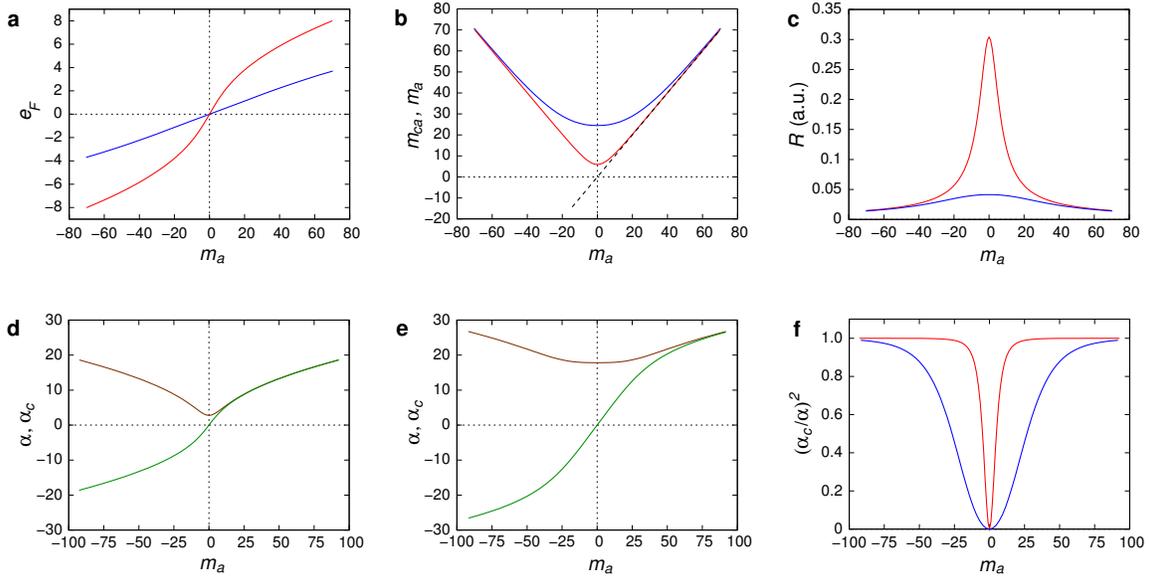

*Figure 1.* Graphene sheet characteristics versus the average normalized density $m_a$ of the charge carriers. (a) Normalized Fermi level $e_F$ showing the much lower value of $|e_F|$ in BLGS (lower curve, for $m_a > 0$) than in SLGS (upper curve, for $m_a > 0$). (b) Normalized density $m_{ca}$ of the current carriers, exhibiting a much greater value (around the Dirac point $m_a = 0$) in BLGS (upper curve) than in SLGS (lower curve) and than $|m_a|$ (dashed curve). (c) Plots of the 'bare' resistance $R \propto 1/m_c$ showing its remarkably larger value (for $|m_a| < 20$) and stronger gate dependence for SLGS (upper curve) than for BLGS (lower curve). (d) [(e)] Plots of $\alpha \equiv \partial m / \partial e_f$ (upper curve) and $\alpha_c \equiv \partial m_c / \partial e_f$ (lower curve) of SLGS [BLGS], showing that $\alpha$ is much greater in BLGS than in SLGS. (f) Plots of $(\alpha_c/\alpha)^2 = (\partial m_c/\partial m)^2$ showing that the interval within which they vary from 0 to 1 is about four times greater in BLGS (lower curve) than in SLGS (upper curve).

These differences between the SLGS and BLGS derive from the fact that there is a much wider range of the density $|m_a|$ around the Dirac point in BLGS than in SLGS for which $m_{ca} > |m_a|$, about 100 against 20 (figure 1 b). Therefore within such a range both electrons and holes are present. From a physical point of view all of this depends in turn on the profound differences existing between the dispersion relationships of the two materials: conic in SLGS and of Mexican hat type in BLGS that, according to (13) [26], has an about flat and wide minimum and, more in general, a much greater $p$ range at the same low energies within which electrons and holes can coexist around the Dirac point.

The proportionality of $<S>$ with respect to $I^2$ and $1/f^\gamma$ is shown by all the experimental results [3-14,25], as well as that with respect to $1/A$ [4,8,13]. However, apart from unknown multiplicative constants, the factor of $<S>$ for which a comparison with experiments is most interesting is $S_a \equiv (1/LW) \int (1 + \beta e_f)[\alpha_c/(\alpha m_c)]^2 P(\eta) d\eta$, which depends on $V_g$ through $m_a$ and which we report versus $m_a$ in Fig 2 in an arbitrary scale.

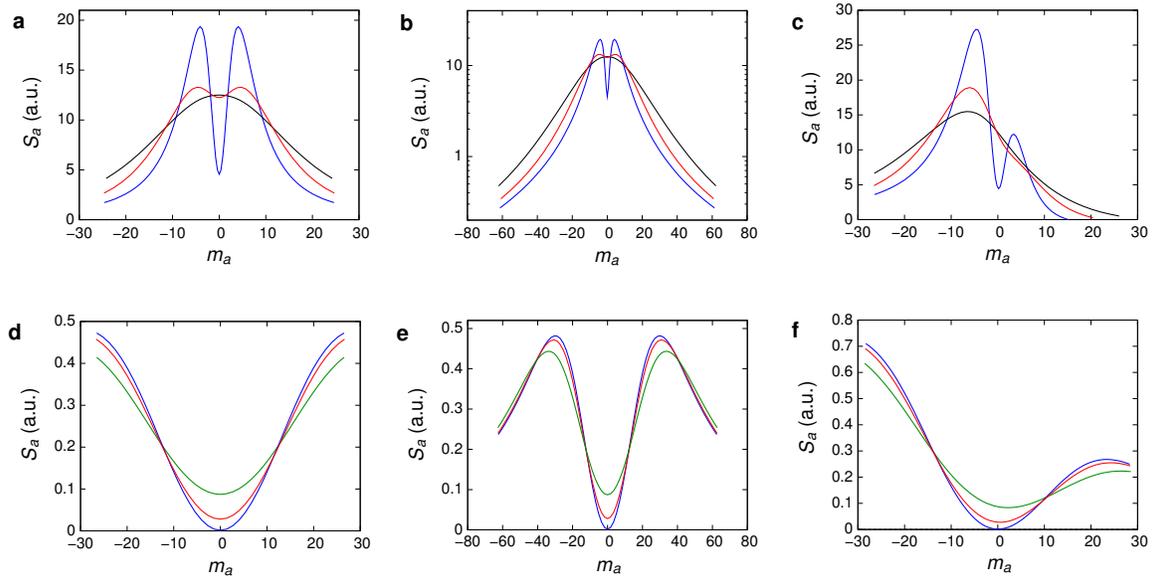

*Figure 2.* (a) and (b) Factor $S_a$ of the SLGS power spectral densities (in arbitrary units (a.u.) for $S_a$) and in different $m_a$ ranges in (a) and (b) for $\sigma_d$ = 1, 3 and 4.5 (curves with largest, intermediate and smallest peaks, respectively). The plots show the switch from the M to the $\Lambda$ shape due to the increase of $\sigma_d$. (c) $S_a$ plots for SLGS showing the effects of the parameter $\beta$ ($\beta = -0.3$) on the evolution of the shape from M to the $\Lambda$ type for $\sigma_d = 3$ and on the shift of the peak with respect to the Dirac point $m_a = 0$. (d) $S_a$ plots for BLGS for $\sigma_d = 1$, 4.5 and 9 (curves with the lowest, intermediate and highest value at $m_a = 0$, respectively) showing a V shape of the *PSD* independent of the above $\sigma_d$ values in a wide $m_a$ range and with values much smaller than for SLGS at $m_a = 0$. (e) $S_a$ plots of BLGS showing that in wide $m_a$ ranges ($|m_a| > 30$) its shape becomes of M type, too. (f) $S_a$ plots of BLGS showing that the parameter $\beta$ ($\beta = -0.3$) does not change the *PSD* V shape and it does not appreciably shift its minimum from the Dirac point.

The plots of figures 2a and 2b for SLGS show, in different $m_a$ ranges, that the behavior of $S_a$ becomes of $\Lambda$ and M type for $\sigma_d > 4.5$ and $\sigma_d \leq 3$, respectively, i.e., as the device inhomogeneity decreases (it becomes null for $\sigma_d = 0$ at $m_a = 0$). We observe a good qualitative agreement, and even quantitative, with the experimental results. In refs 3-5, 11 and 25 we have a $\Lambda$ shape in about a *PSD* decade in the $|m_a|$ ranges (0, 40) and (0, 80), and in the range of (80, 290) K for the temperature, in agreement with figure 2b. In refs 5-8, 12 we have an M shape; in particular in the devices with suspended graphene [5,8] the spectrum maxima are within the range (0, 4) of $|m_a|$ given by figure 2a, whereas wider intervals of $|V_g| \approx |m_a|$ exist in the devices [6-8] that are not suspended because the charge stored by the substrate traps, which depends on $V_g$, partly screens that of graphene. In ref 7 the M shape is obtained versus the voltages of both the back gate and an electrolyte top gate. Figure 2c shows how the parameter $\beta$ ($\beta = -0.3$), depending on the distribution of the trap energy levels, shifts the *PSD* peak away from the Dirac point where the resistance reaches its maximum [13] and can change the M shape into a $\Lambda$ shape, e.g., for $\sigma_d = 3$.

The BLGS spectra $S_a$ are given in figures 2d, 2e and 2f, and they are compared with those for SLGS in the same $|m_a|$ scales. We observe three significant differences. In BLGS the *PSD* has a V shape that practically does not depend on $\sigma_d$ ($\sigma_d$ = 1, 4.5 and 9), i.e., on the device inhomogeneity, whereas in SLGS it is very sensitive to $\sigma_d$, transitioning from $\Lambda$ to M as $\sigma_d$ decreases, and its value is much smaller in the first than in the second case. For $|n| > 2 \times 10^{12}$ cm$^{-2}$, as shown in figure 2e, the BLGS *PSD* decreases as $|n|$ increases and it also becomes of M type, whereas for the SLGS this happens above $|n| > 2.5 \times 10^{11}$ cm$^{-2}$. The plots in figure 2f show the effect of a value of $\beta = -0.3$ that maintains the V shape and its minimum at about $m_a = 0$. Such differences are due to the larger values in BLGS of $m_{ca}$ (figure 1b) and of $\alpha$ (figures 1d and 1e) which, according to (13), lower the *PSD* and the standard deviation $\sigma = \sigma_d / \alpha$, that is the inhomogeneity effects, as well as due to the related wider interval of $m_a$ (figure 1f) in which $(\alpha_c/\alpha)^2$ varies between 0 and 1. These differences derive from the fact that there is a much wider range of the density $m_a$ in BLGS than in SLGS within which we can have bipolar conduction, that is both electrons and holes are present and are affected in opposite ways by the charge fluctuations in the traps, thereby decreasing the noise. As previously highlighted, this is the result of the profound differences existing between the dispersion relationships of the two materials. The V shape for BLGS agrees well with most of the experimental results [4-7,9-11, 14], whereas the M shape has been found for $|n| > 2 \times 10^{12}$ cm$^{-2}$ in ref 5, in accordance with figure 2e, and in ref 7. The V shape has been found also in multilayer devices [25].

6.2. *Ribbon devices*

For a comparison with the experimental results relative to graphene ribbons, let us consider a SLGR with $W$=30 nm and three BLGRs with $W$=30, 40 and 80 nm (dealt with in ref 13), for which in $\sigma = \sigma_d/\alpha$ we use the same fitting parameter $\sigma_d$ because the devices are fabricated with the same process. The computation is performed for metallic and semiconducting ribbons as well as for their mixed behavior. Here let us consider metallic 30 nm wide BLGR and SLGR for $\sigma_d = 9$, for which the plots of $e_F$, $m_a$, $m_{ca}$, $\alpha$ and $\alpha_c$ versus $m_a$ shown in figure E1 are completely analogous to those of the sheets of figure 1.

The material inhomogeneities generate a percolative structure for carrier transport which would not allow averaging the device resistance $R = L/[q\mu n_0 m_c(W - W^*)]$ according to the previous procedure (Appendix D). This, however, becomes possible for SLGR, in which the screening length is greater than in BLGR (Appendix C), and for the smallest ribbon width ($W - W^*$) = 14 nm: these facts and the inhomogeneities with disorder length scale of about 30 nm [24], can generate ribbon segments in series to each other, so that the average resistance-like quantity $R_a(\sigma_d) \equiv q\mu n_0 <R> = [L/(W-W^*)]\int m_c^{-1} P(\eta)d\eta$ can be exploited in the smallest SLGR ribbon, unlike in the BLGR ones, for which, instead, the non averaged value $R_a(0) = q\mu n_0 R = L/[m_c(W-W^*)]$ is employed. For the *PSD*, an element of interest to be compared with the experiments, as in sheet devices, is $S_a \equiv [1/L(W-W^*)]\int [\alpha_c/(\alpha m_c)]^2 P(\eta)d\eta$ (for $\beta = 0$).

The plots of $R_a$ and $S_a$ versus $m_a$ evaluated for $W = 30$ nm and for $\sigma_d = 9$ and given in figure 3 for a metallic behavior show a good agreement with the experimental results of ref 13. In the SLGR the 'resistance' $R_a(9)$ has a larger value (for $|m_a| < 40$) and a stronger gate dependence than the bare one $R_a(0)$ in BLGR (figure 3a). The resistance and the *PSD* have the same $\Lambda$ shape in SLGR (figure 3b), whereas the *PSD* in the BLGR becomes of V type (figure 3c) and, at the Dirac point, its value is much smaller than for the BLGR. Such differences have the same physical origin as in sheet devices.

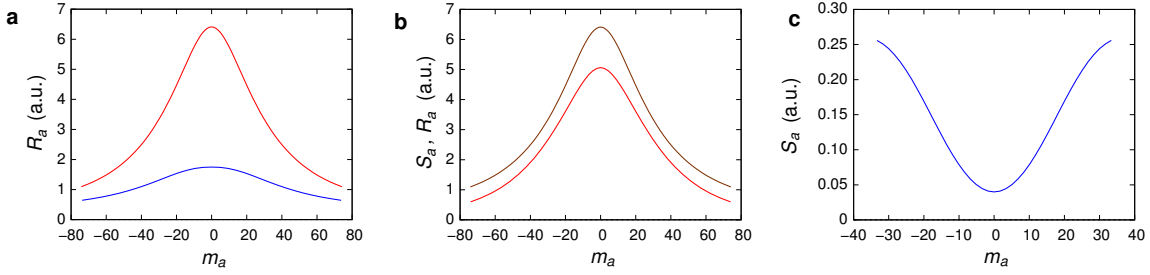

*Figure 3*. Comparison of characteristics versus $m_a$ of single-layer and bilayer metallic graphene ribbons of width $W = 30$ nm, for $\sigma_d = 9$. (a) Comparison of the SLGR average 'resistance' $R_a(9)$ (upper curve) with the BLGR bare one $R_a(0)$ (lower curve), showing a larger value (for $|m_a| < 40$) and a stronger gate dependence for SLGR than for BLGR. (b) Plots of $R_a$ (upper curve) and $S_a$ (lower curve) of the SLGR showing their equal $\Lambda$ shape. (c) $S_a$ plot of the BLGR showing the large difference in shape, this time of V type, and in term of value, this time much lower, in comparison to $S_a$ of $\Lambda$ type of panel (b) of the SLGR.

A more quantitative comparison arises from the noise factors $\psi$ shown in figure 4 for metallic, semiconducting and mixed ribbons.

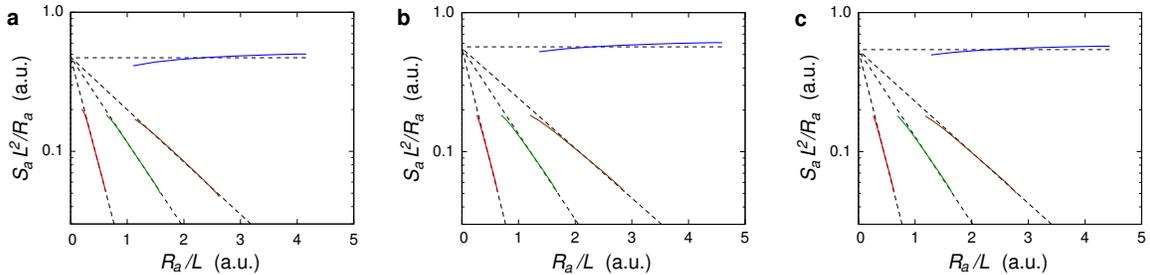

*Figure 4.* (a) The topmost curve corresponds to the noise factor $\psi \equiv S_a L^2 / R_a(\sigma_d)$ of a 30 nm wide metallic SLGR versus the average 'resistance' per unit of length $R_a(\sigma_d)/L$, both in arbitrary units (a.u.). The other curves, whose slope increases with $W$, show the behavior of the noise factor $\psi \equiv S_a L^2 / R_a(0)$ versus the bare 'resistance' per unit of length $R_a(0)/L$ for three metallic BLGR with widths $W = 30, 40$

and 80 nm. All the plots are computed for the same $\sigma_d = 29$. For SLGR the factor $\psi \approx \psi_0$ is nearly constant in a resistance interval characterized by a ratio $\rho_M \approx 3.8$ between its boundaries, while for BLGR, in a logarithmic scale for the ordinate axis, $\psi$ is linear in resistance intervals with ratios $\rho_B = 2.3 \div 2.5$ between the boundaries; its asymptotic limit for null resistance equals $\psi_0$ and its minimum value is more than a decade lower than $\psi_0$ itself. (b) As in (a) for a semiconducting behavior, with $\sigma_d = 26$, $\rho_M \approx 3.3$ and $\rho_B = 1.9 \div 2.2$. (c) Same as in (a) and (b) but for a mixed ribbon, with $\sigma_d = 27$, $\rho_M \approx 3.4$ and $\rho_B = 2.0 \div 2.1$, for which, since we have two possible cases for the semiconducting behavior and one for the metallic behavior, we consider weights of 2/3 and 1/3, respectively.

The upper plots show the noise factor $\psi \equiv S_a L^2 / R_a(\sigma_d) = \int [\alpha_c / (\alpha m_c)]^2 P(\eta) d\eta / \int m_c^{-1} P(\eta) d\eta$ of a 30 nm wide SLGR versus the average 'resistance' per unit of length $R_a(\sigma_d) / L = (W - W^*)^{-1} \int m_c^{-1} P(\eta) d\eta$, while the inclined plots, whose slope increases with $W$, report the behavior of $\psi \equiv S_a L^2 / R_a(0) = m_c \int [\alpha_c / (\alpha m_c)]^2 P(\eta) d\eta$ versus $R_a(0) / L = [m_c(W - W^*)]^{-1}$ for three BLGRs with widths $W = 30$, 40 and 80 nm, for which, instead, we consider the bare 'resistance' $R_a(0)$. The plots of Fig. 4a, 4b and 4c are relative to the metallic, semiconducting, and mixed behavior of the ribbon, respectively, and the corresponding values of $\sigma_d$ are 29, 26 and 27. We substantially obtain the same results in all three cases. The SLGR noise factor $\psi \approx \psi_0$ is nearly constant within resistance ranges characterized by a ratio $\rho_M = 3.3 \div 3.8$ between their boundaries. The BLGR factors $\psi$, if represented in a semilogarithmic scale on the vertical axis as shown in figure 4, are linear in the ranges with boundary ratios $\rho_B = 1.9 \div 2.5$, their asymptotic values for null resistance equals $\psi_0$, and their minimum value is a decade lower than $\psi_0$ itself. The largest linear intervals, i.e., the largest values of $\rho_M$ and $\rho_B$, are obtained for the metallic ribbon. All these results agree well with the experimental ones on the same devices of ref 13 in which we have $\rho_M = 3.0 \div 3.5$ for SLGR and $\rho_B = 1.6 \div 2.2$ for BLGR, and of the same order as the difference between $\psi_0$ for SLGR and the minima for BLGR. In particular, the result that $\psi \approx \psi_0$ is nearly constant for SLGRs does not mean that the spectrum obeys the empirical relationship of refs 23 and 13 because it is not true that $R_a \propto 1/N = 1/(An_0 m_a)$.

## Conclusions

In conclusion a comprehensive and unitary model has been presented, which explains most of the 1/$f$ noise experimental properties of graphene devices, sheds light on their transport mechanism, can be utilized as a tool to characterize nanostructures, can improve their technological applications and is open to further developments in other fields. Indeed, the proposed approach can be extended to the non ohmic conduction, to asymmetric conduction and valence bands, to multilayers [11,25], to twisted bilayer [32] and chemically functionalized [33,34] graphene, to carbon nanotube devices [35] and to new quantum materials such as topological insulators [36,37].

## Acknowledgments

The author whishes to thank M. Macucci, P. Marconcini and G. Fiori for useful discussions, P. Marconcini also for help in using some computing and plotting tools and M. Macucci also for the English revision.

The work, in part, was supported by the European Union under Contract No. 215752 GRAND (GRAphene-based Nanoelectronic Devices). The author has no competing financial interest.

## Appendix A

The second term of equation (2) vanishes in several cases of interest for the applications. Of course it is null when $S_R = 0$, i.e., when the electrodes completely enclose $\Omega$ (and the Ramo-Shockley theorem is applicable as a particular case of the electrokinematics theorem [38]). Also for frequencies $f = \omega/(2\pi) \ll 1/(2\pi t_j)$, the second term, due to the time derivatives, provides no contribution to the power spectral density, because, according to the Fourier transform, it derives from integrals such as $\int_0^{t_j} \exp(-\omega t)(\partial Q/\partial t)dt \approx Q(t_j) - Q(0)$, in which $Q(t_j) = Q(0) = 0$. Moreover, owing to the screening effect of the other carriers, such a term gives a negligible contribution even if such a condition on the frequency is not satisfied, for instance when the $j$th carrier is stored for a long time $\tau_j$ in a trap at $r_t$ during its travel between electrodes so that it can be $t_j = t'_j + \tau_j + t''_j > 1/\omega$, being $t'_j \ll 1/\omega$ and $t''_j \ll 1/\omega$ the travel times between trap and electrodes. In reality, due to the arbitrariness of $F$ and of $S_R$ we can choose them in such a way that $F$ is tangent to $S_R$ in any point, so that the integral of $(\partial A_{0j}/\partial t)F$ vanishes. Furthermore, in the zone 'around' $r_t$, let us choose a close surface consisting of a part $S_R^{'}$ of $S_R$ bounded by a line $l_R$ where $\mathbf{E}_j \approx 0$ and by a surface $S_T$ external to the device that starts from $l_R$ and on which $\mathbf{E}_j \approx 0$, too. These conditions can be satisfied if the screening length is much smaller than the linear dimensions of $S_R$, as it happens in graphene devices (Appendix C), so that, moreover, $\Phi$ becomes nearly constant on $S_R^{'}$. Therefore, after the Fourier transform, we have to perform the integral $\int_{S_{R'}} \Phi \varepsilon \mathbf{E}_j \cdot d\mathbf{S} \approx \Phi \int_{S_{R'}} \varepsilon \mathbf{E}_j \cdot d\mathbf{S}$ that, according to Gauss's theorem applied to the volume enclosed by $S_R^{'} + S_T$ (which does not contain charge), vanishes or becomes negligible, in particular with respect to the contribution due to variation induced by the trap charge $j$th on $\mathbf{J}_c$ in the r.h.s. of equation (2). Moreover, in a more exact way, if the device is a parallelepiped, as in our case, or a cylinder, as in the case carbon nanotubes, parallel to the $x$ axis and the bases separated by a distance $L$ coincide with the electrodes ($S_E$), one can choose $\Phi = x/L$, (i.e., $\mathbf{F} \equiv F_x \mathbf{i} \equiv -(1/L)\mathbf{i}$) so that we have $\int_{S_{R'}} \Phi \varepsilon \mathbf{E}_j \cdot d\mathbf{S} = (x_t/L)\int_{S_{R'}} \varepsilon \mathbf{E}_j \cdot d\mathbf{S} + (1/L)\int_{S_{R'}} (x - x_t)\varepsilon \mathbf{E}_j \cdot d\mathbf{S}$ where the second integral, according to the preceding procedure, is null, and also the third integral is null as a consequence of the symmetry of $\mathbf{E}_j$ along $x$ around the trap abscissa $x_t$. Therefore equation (2) it becomes equation (3).

**Appendix B**

In the case of different mobilities $\mu_n$ and $\mu_p$ of electrons and holes, respectively, the drift current density becomes $J_{cx} = q(\mu_n n_n + \mu_p n_p)E$ and equation (5) is replaced by

$$i = -\frac{1}{L}\int_A q(\mu n_c + \mu_d n)E dx dy, \quad (B.1)$$

where $\mu \equiv (\mu_n + \mu_p)/2$ and $\mu_d \equiv (\mu_n - \mu_p)/2$. The current $i$ depends on $\mu_n$ and $\mu_p$ that in their turn depend on the carrier densities $n_n$ and $n_p$, respectively [39].

In this case the current fluctuations are given by the equation

$$\frac{\Delta i}{I} = \frac{1}{A}[\int_A \frac{\mu \Delta n_c + \mu_d \Delta n}{\mu n_c + \mu_d n}dxdy + \int_A \frac{\Delta E}{E}dxdy + \int_A \frac{n_n \Delta \mu_n + n_p \Delta \mu_p}{\mu n_c + \mu_d n}dxdy], \quad (B.2)$$

in which the various quantities $n_c$, $E$, $\mu$ and $\mu_c$ depend on $n_n$ and $n_p$ and, in general, their fluctuations can be generated by either the same or different noise sources, and thus they can be either correlated or uncorrelated. Therefore the model of the current fluctuations can become complex. However, even if a mixture of fluctuating mechanisms with different relative weight is possible, according to most experimental results, the prevalence of the trapping-detrapping process as source of 1/$f$ noise in graphene seems possible also for $\mu_n \neq \mu_p$ and we can derive the same conclusions deduced from the equation (6). In particular from equation (B.2), we obtain the equation

$$\frac{\Delta i}{I} = -\frac{1}{A} \frac{\mu(a_c/a) + \mu_d}{\mu n_c + \mu_d n} \Delta \chi \approx -\frac{1}{A}(\frac{a_c}{a n_c})\Delta \chi \,, \text{(B.3)}$$

where the last term holds if we take into account that it is $|\mu_d/\mu| < 1$ and $|n/n_c| < 1$ (and possibly $|\mu_d/\mu| \ll 1$ and $|n/n_c| \ll 1$), and $|a_c/a| \approx 1$, at least away from the Dirac point $n=0$ (see figure 1). Therefore equation (8), to the first order, can be exploited also for $\mu_n \neq \mu_p$.

**Appendix C**

The variations $\Delta n$ and $\Delta A_0$ generated by the trap charge fluctuations $-q\Delta\chi$ substantially occur in the $x\, y$ plane, rather than in the perpendicular direction $z$, so that, in a simplified model of the screening length, we can assume a uniform carrier volume density $n/c_0 = n_0 m(e_f + q\Delta A_0/k_B T)/c_0$ in graphene along the $z$ axis, with $c_0 = 3.35\ \text{Å}$ [26] and $c_0 = 0.8 \div 1\ \text{Å}$ for bilayer and single-layer graphene, respectively. This is equivalent, in the first case, to neglecting the difference between the densities $n_1$ and $n_2$ of the two layers, being $n_1 + n_2 = n$. Therefore, the Poisson equation for the potential variation $\Delta A_0$ in the graphene channel reads

$$\nabla^2(\Delta A_0) = \frac{q}{\varepsilon_r \varepsilon_0}\{\frac{n_0}{c_0}[m(e_f + q\Delta A_0/k_B T) - m(e_f)] + \Delta \chi\ \delta(\mathbf{r}\text{-}\mathbf{r}_t)\}, \text{(C.1)}$$

where $\varepsilon_0$ is the electric permittivity of free space, $\varepsilon_r = 1 \div 3.3$ is the graphene dielectric constant. At the first order in $\Delta A_0$, equation (B.1) becomes

$$\nabla^2(\Delta A_0) = \frac{\Delta A_0}{\lambda_s^2} + \frac{q}{\varepsilon_r \varepsilon_0} \Delta \chi \, \delta(\mathbf{r}-\mathbf{r}_t)\} \,, \tag{C.2}$$

where

$$\lambda_s \equiv \frac{1}{q}\left(\frac{\varepsilon_r \varepsilon_0 c_0 k_B T}{\alpha n_0}\right)^{1/2} \,, \tag{C.3}$$

is the scale factor of $\Delta A_0$ on the graphene plane (for the definition of the quantities $n_0$, $m$ and $\alpha$ one sees section 4.1.). Since the lowest value of $\alpha$, at $e_f = 0$, is 2.8 and 17.8 for the single layer and bilayer graphene, respectively (figures 1d and 1e), for $n_0 = 7.7 \times 10^{10}$ cm$^{-2}$ and $\varepsilon_r = 3.3$ of graphite [40], we obtain $\lambda_s < 4.7$ Å and $\lambda_s < 3.4$ for single layer and bilayer devices, respectively, that agree with the values of 3.8 and 5 Å quoted as graphite screening length in the c-axis in refs 13 and 21.

**Appendix D**

For an additive quantity $H(\varepsilon_f)$, such as the carrier densities and the *PSD*, the average value over the surface *A* becomes

$$<H(\varepsilon_F)> \equiv \frac{1}{A}\delta A \sum_{j=1}^{M'} H(\varepsilon_{fj}) n_j = \frac{1}{A}(\delta A \sum_{j=1}^{M'} n_j) \sum_{j=1}^{M'}[H(\varepsilon_{fj})(n_j/\sum_{j=1}^{M'} n_j)] =$$

$$\sum_{j=1}^{M'} H(\varepsilon_{fj}) P'(\varepsilon_{fj}) = \int H(\varepsilon_F + \varepsilon_{f\delta}) P(\varepsilon_{f\delta}) d\varepsilon_{f\delta} \,, \tag{D.1}$$

where $\delta A$ is the smallest surface of *A* over which the Fermi level $\varepsilon_{fj}$ is constant, $n_j$ is the number of surfaces $\delta A$ having the same $\varepsilon_f = \varepsilon_{fj}$, *M'* is the number of distinct $\varepsilon_{fj}$, $\sum_{j=1}^{M'} n_j$ is the number of $\delta A$ into which *A* can be subdivided, $P'(\varepsilon_{fj})$ is the probability of $\varepsilon_{fj}$, whereas, by setting $\varepsilon_f = \varepsilon_F + \varepsilon_{f\delta}$ with $\varepsilon_F \equiv <\varepsilon_f>$, $P(\varepsilon_{f\delta})$ is the distribution function of $\varepsilon_{f\delta}$. Then, from (D.1), the noise power spectral density is given by (10).

**Appendix E**

For the ribbons, from (21) we have $e_{nh}|_{e_x=0} \geq e_f + \vartheta$ for $u \geq [2\sqrt{\gamma_1(e_f + \vartheta)/(k_B T)}/(3e_w) + 1]$ and $u \geq [2(e_f + \vartheta)/(3e_w) + 1]$ in the BLGR and SLGR cases, respectively. Therefore, e.g., for $\vartheta \geq 8$, the contribution of the energy sub-bands with $h \geq u$ becomes negligible. Moreover, since $l$ is an integer in the range $(0, M+2)$, we also have the condition $-2(M+2) < \pm s < (M+2)$ that requires, in the more restrictive case, $u < (M+5)/3$ and $u < (2M+5)/3$ for metallic and semiconducting nanoribbons, respectively.

For the ribbons in figure E.1 we also show the plots of $e_F$, $m_{ca}$, $m_a$, $\alpha$, $\alpha_c$ and $(\alpha_c/\alpha)^2$ for $\sigma_d = 9$ of a 30 nm wide metallic ribbon that are analogous to the corresponding ones of the 2D graphene sheets of figure 1.

Moreover, in order to obtain $\Delta/(k_B T) \approx cn_0 m/(k_B T)$ through $e_f$, we approximate $m$ given by the equation (11) by means of $m = a' e_f + b' e_f^3$. As shown in figure D.1a, we obtain, for the BLGS, a good accuracy with $a' = 18$ and $b' = 0.10$. The same result holds true for the semiconducting BLGR, while the metallic BLGR requires $b' = 0.14$.

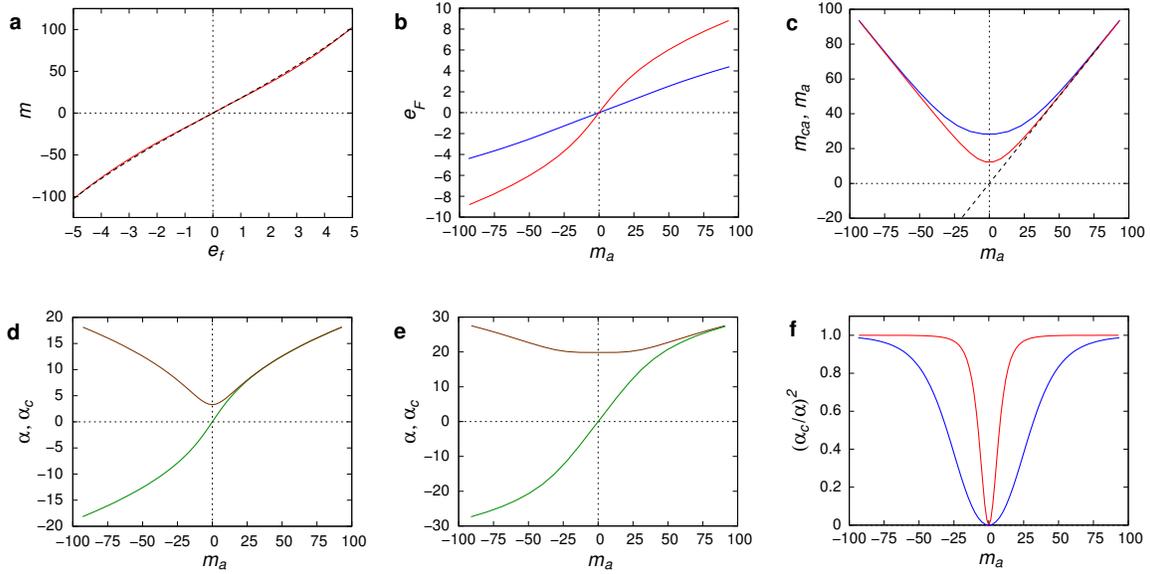

*Figure* E1. (a) Comparison of the plots of $m$ versus the Fermi level $e_f$ obtained from equation (11) (solid line) and from its approximation $m = a' e_f + b' e_f^3$ (dashed line). (b), (c), (d), (e) and (f) Plots of $e_F$, $m_{ca}$,

$m_a$, $\alpha$, $\alpha_c$ and $(\alpha_c/\alpha)^2$ versus $m_a$, respectively, for metallic 30 nm wide BLGR and SLGR and $\sigma_d = 9$.

**References**

1. K. S. Noveselov, A. K. Geim, S. V. Morozov, D. Jiang, Y. Zhang, S. V. Dubonos, I. V. Grigorieva, A. A. Firsov, Scince **306**, 666 (2004).
2. K. S. Noveselov, A. K. Geim, S. V. Morozov, D. Jiang, M. I. Katsnelson, I. V. Grigorieva, S. V. Dubonos, A. A. Firsov, Nature **438**, 197 (2005).
3. A. N. Pal, A. A. Bol, A. Ghosh, Appl. Phys. Lett. **97**, 133504 (2010).
4. S. Rumyantsev, G. Liu, W. Stillman, M. Shur, A. A. Balandin, J. Phys.:Condens. Matter **22**, 395302 (2010).
5. A. N. Pal, S. Ghatak, V. Kochat, E. S. Sneha, A. Sampathkumar, S. Raghavan, A. Ghosh, ACS Nano **5**, 2075 (2011).
6. G. Xu, C.M. Torres. Y. Zhang, F. Liu, E. B. Song, M. Wang, Y. Zhou, C. Zeng, K. L. Wang, Nano Lett. **10**, 3312 (2010).
7. I. Heller, S. Chatoor, J. Männik, M. A. G. Zevenbergen, J. B. Oostinga, A.F. Morpurgo, C. Dekker, S. G., Lemay, Nano Lett. **10**, 1563 (2010).
8. Y. Zhang, E. E. Mendez, X. Du, ACS Nano **10**, 8124 (2011).
9. A. N. Pal, A. Ghosh, Phys. Rev. Lett. **102,** 126805 (2009).
10. Q. Shao, G. Liu, D. Teweldebrhan, A. A. Balandin, S. Rumyantsev, M. Shur, D. Yan, IEEE Electron Device Lett. (IEDL) **30**, 288 (2009).
11. A. N. Pal, A. Ghosh, Appl. Phys. Lett. **95**, 082105 (2009).
12. A. A. Kaverzin, A. S. Mayorov, A. Shytov, D. W. Horsel, Phys. Rev. **B 85**, 075435 (2012).
13. Y.M. Lin, P. Avouris, Nano Lett. **8**, 2119 (2008).
14. M. Z. Hossain, S. Rumyantsev, M. S. Shur, A. A. Balandin, App. Phys Lett. **102**, 153512 (2013).
15. B. Pellegrini, Microelectronics Reliability **40**, 1775 (2000).
16. B. Pellegrini, Phys. Rev. **B 34**, 5921 (1986).
17. B. Pellegrini, Rivista del Nuovo Cimento **15 D**, 855 (1993).
18. S. Feng, P.A. Lee, A. D. Stone, Phys. Rev. Lett. **56**, 1960 (1986).
19. S. Ramo, Proc. IRE **27**, 584 (1939).
20. W. Shockley, App. Phys. **9**, 635 (1938).
21. D. P. DiVincenzo, E. J. Mele, Phis. Rev.**B 29**, 1685 (1984).
22. S. J. Machlup, Appl. Phys. **25**, 341 (1954).

23. F. N. Hooge, Phys. Lett. **29 A**, 139 (1969).
24. J. Martin, N. Akerman, G. Ulbricht, T. Lohmann, J. H. Smet, K. von Klitzing, A. Yacoby, Nature Phys. **5**, 144 (2008).
25. G. Liu, S. Rumyantsev, M. S. Shur, A. A. Balandin Appl. Phys. **102**, 093111 (2013).
26. E. McCann, Phys, Rev. **B 74**, 161403 (2006).
27. A. V. Rozhkov, S. Savel's, F. Nori, Phys. Rev. **B 79**, 125420 (2009).
28. P. Marconcini, and M. Macucci, Rivista del Nuovo Cimento **34**, 489 (2011).
29. M. Fagotti, C. Bonati, D. Logoteca, P. Marconcini, and M. Macucci, Phys. Rev. **B 83**, 241406 (2011).
30. Y-W. Son, M. L. Cohen, S. G. Louie, Phys. Rev. Lett. **97**, 216803 (2006).
31. M. Y. Han, B. Özyilmaz, Y. Zhang, and P. Kim, Phys. Rev. Lett. **98**, 206805 (2007).
32. P. Moon, M. Koshino, Phys. Rev. **B 85**, 195458 (2012).
33. G. Fiori, S. Lebegue, A. Betti, P. Michetti, M. Klintenberg, Phys. Rev. **B 82**, 153404 (2010).
34. V. J. Surya, K. Iyakutti, H. Mizuseki, Y. Kawazoe, IEEE Trans. Nanotech. **11**, 534 (2012).
35. Y. M. Lin, J. Appenzeller, J. Knoch, Z. Chen, P. Avouris, Nano Lett. **6**, 930 (2006).
36. M. Z. Hossain, S. L. Rumyantsev, K. M. F. Shahil, D. Teweldebrhan, M. Shur, A. A. Balandin, ACS Nano **4**, 2657 (2011).
37. J. E. Moore, Nature **464,** 164 (2010).
38. A. Benali, L. Traversa, G. Albareda, M. Aghoutane, X. Oriols, App. Phys Lett. **102**, 153506(2013).
39. W. Zhu, V. Perebeinos, M. Freitag, and P. Avouris, Phys. Rev. **B 90**, 235402 (2009).
40. E. K. Yu, D. A. Stewart, S. Tiwari, Phys. Rev. **B 77**, 195406 (2008).